
\magnification=1200
\voffset=0.0 true in
\hoffset=0.0 true in
\vsize=7.5in
\hsize=5.6in
\tolerance 10000
\newcount \pageit
\footline={\hss\tenrm\ifnum\pageit=0 \hfill \else \number\pageno \fi\hss}
\pageit=0
\pageno=0
\baselineskip 12pt plus 1pt minus 1pt
\centerline{{\bf A COMMENT ON ENTROPY AND AREA}\footnote{*}
{\rm This work is supported in part by funds provided by the U. S. Department
of Energy (D.O.E.) under contract \#DE--FG05--90ER40559.}}
\vskip 24pt
\centerline{D.~Kabat and M.~J.~Strassler}
\vskip 12pt
\centerline{\it Dept.~of Physics and Astronomy}
\centerline{\it Rutgers University}
\centerline{\it Piscataway, NJ 08855--0849}
\vskip 12pt
\centerline{\it kabat, strasslr@physics.rutgers.edu}
\vskip 0.9truein
\vskip 0.9truein
\centerline{\bf ABSTRACT}
\bigskip
\leftskip = 0.7truein
\rightskip = 0.7truein
\noindent
For an arbitrary quantum field in flat space with a planar
boundary, an entropy of entanglement, associated with correlations across
the boundary, is present when the field is in its vacuum state.
The vacuum state of the same quantum field appears thermal
in Rindler space, with an associated thermal entropy.  We show that the
density matrices describing the two situations are identical, and therefore
that the two entropies are equal.
We comment on the generality and significance of this result, and make use
of it in analyzing the area and cutoff dependence of the entropy.
The equivalence of the density matrices leads us to speculate that a
planar boundary in Minkowski space
has a classical entropy given by the Bekenstein--Hawking formula.

\leftskip = 0.0truein
\rightskip = 0.0truein
\vfill
\vskip -28pt
\noindent hep-th/9401125
\smallskip
\par
\noindent RU--94--10  \hfill  January 1994
\eject

\pageit=1
\baselineskip 24pt plus 2pt minus 2pt

\bigskip
\goodbreak
\noindent{\bf INTRODUCTION}
\nobreak
\medskip
\nobreak
In a recent paper Srednicki$^1$ considered the entropy of entanglement
of a quantum field.  Taking a free
scalar field in its ground state, he calculated the density matrix
which describes the state of the field outside an imaginary sphere when
one chooses not to make measurements in the sphere's interior.  He
found numerically that the corresponding entropy scales as the surface
area of the sphere. This entropy was originally studied by Bombelli
{\it et. al.}$^2$ as a quantum correction to the entropy of a black
hole.

't Hooft has studied a seemingly different source of quantum corrections
to the entropy of a black hole arising
from scalar fields propagating in the region just outside the horizon$^3$.
For very massive black holes this region may be approximated as flat Rindler
space, in which quantum fields are thermally excited
at the Hawking--Unruh temperature and carry a corresponding thermal entropy.
This entropy was likewise found to scale as the area of the horizon.

The purpose of this comment is to point out that, for a sphere of infinite
radius, the density matrix studied by Srednicki is equal to the thermal
density matrix describing the fields studied by 't Hooft.
(This is a special case of a more general result proven
in the next section.)
It follows that the two entropies are identical in this limit
and should be thought of as the same quantity
measured by different classes of observers. The possibility of this
equivalence was suggested by  Bombelli {\it et.~al.}$^2$.
Unfortunately the entropy is
divergent; our discussion is therefore limited to its
dependence on the area of the boundary and the degree of its divergence.
Except in 1+1 dimensions, the coefficient of the divergence is
non-universal.  We will explain below that, as
a result of subtleties involving regulators, Srednicki's
actual computation$^1$
cannot be quantitatively compared with any of 't Hooft's results.$^3$

Other groups are investigating these issues and have reached similar
conclusions$^{4,5,6}$.

\bigskip
\goodbreak
\noindent{\bf PROOF OF EQUIVALENCE}
\nobreak
\medskip
\nobreak
In this section we show that for the vacuum state of a
field theory,
the density matrix found by tracing over
the fields in half of space is equal to the thermal
density matrix describing the fields in Rindler space.
This result holds for a wide class of field theories; we will
discuss its generality in the next section.

Consider a field $\phi(t,x,y_1,\ldots,y_{d_\perp})$ in a $(d_\perp+1)+1$
dimensional Minkowski space.  We will describe the field configuration
$\phi(t,x,y_i)$ as $\phi_R(t,x,y_i)$ for $x>0$ and as
$\phi_L(t,x,y_i)$ for $x<0$; in subsequent expressions we suppress
the dependence on the variables $y_i$.
At a fixed time $t_0$ take the field to
be in its ground state, with density matrix $\vert 0><0\vert$, and form the
density matrix $\rho_R$ describing the state
of the field $\phi_R(t_0,x)$ for $x>0$ by tracing over all degrees of freedom
$\phi_L(t_0,x)$
located at $x<0$.
$$<\phi'_R|\rho_R|\phi^{\prime\prime}_R> = \int {\cal D} \phi_L \,
        <\phi_L\phi'_R | 0> \, <0|\phi_L\phi^{\prime\prime}_R> \,  \eqno(1)$$
The entropy of entanglement is defined to be
$$S = - \int {\cal D} \phi_R <\phi_R | \rho_R \log \rho_R | \phi_R>\, .
\eqno(2)$$

To show that this entropy is thermal, we
introduce a Euclidean path integral representation for the density
matrix (1).  First, we use a Euclidean functional integral to
generate projections onto the vacuum.
$$<0|\phi'_L\phi'_R> = \int {\cal D}\phi \,\,{\rm exp} \Big[- \int_0^\infty
d \tau \int dx\ d^{d_\perp}y \, {\cal L} \Big]$$
The integral is over all fields $\phi(\tau>0,x)$ that at $\tau = 0$
take on the value $\phi'_L(x)$ for $x<0$ and $\phi'_R(x)$ for $x>0$.
Putting two of these functional integrals together and integrating over
$\phi'_L(x)$ leads to a representation of the density matrix element (1) as a
single Euclidean functional integral on a space with a
cut at the set of points $\lbrace \tau=0, x>0 \rbrace$,
subject to the boundary condition that
$\phi(\tau=+\epsilon,x>0) = \phi^{\prime\prime}_R(x)$ just above the cut and
$\phi(\tau=-\epsilon,x>0) =\phi'_R(x)$ just below it,
$$\eqalign{
<\phi'_R|\rho_R|\phi^{\prime\prime}_R> = C \ \int {\cal D}\phi \,\,
& e^{- S}\ \delta[\phi_R(\tau=-\epsilon)-\phi'_R]\
\delta[\phi_R(\tau=+\epsilon)-\phi^{\prime\prime}_R]\cr
&\qquad{\rm exp} \left\lbrace + \int_{- \epsilon}^{+ \epsilon} d\tau
\int_0^\infty dx \int d^{d_\perp} y \, {\cal L} \right\rbrace \, .
\cr}$$
Here $\epsilon$ is infinitesimal and $C$ is chosen to normalize
${\rm Tr} \, \rho_R = 1$.

Now think of this path integral as generating time evolution under
some Hamiltonian.  In this geometry, instead of taking the usual Hamiltonian
and time slices at fixed $\tau$, it is natural to use angular
quantization$^7$.
Introduce the Euclidean Rindler Hamiltonian $H_R$ which is the generator of
rotations in the $\tau$--$x$ plane (in Minkowski space it is the
Lorentz boost generator).  The above path integral is then seen to be
$<\phi'_R|{\rm exp}\left(-2\pi H_R\right)|\phi^{\prime\prime}_R>$, so
$\rho_R = e^{-2 \pi H_R}\,.$\footnote{*}{This is nothing more than
the standard Euclidean demonstration that
the Minkowski vacuum looks thermal to a Rindler observer.}
This completes the formal proof that the density matrix in
half of Minkowski
space is a thermal ensemble with respect to the Rindler Hamiltonian,
with an inverse temperature $\beta = 2\pi$.  \footnote{**}{The statement that
the density matrix is thermal requires clarification.  Any
density matrix has a
well defined logarithm, so $\rho = e^{-H}$ for some $H$.  This case is
special in that $H$ is the simple local
operator $H_R$.}

\bigskip
\goodbreak
\noindent{\bf DISCUSSION}
\nobreak
\medskip
\nobreak
This equivalence is very satisfying: the two
seemingly different density matrices are actually
the same density matrix interpreted
by different observers.  An inertial observer who chooses not to make
measurements on the field at $x<0$ obtains exactly the same results for any
experiment as a Rindler observer who is prevented by a horizon
from making measurements at $x<0$.  In particular the two entropies are
formally identical, which is a great technical simplification, since the
entropy
of entanglement defined in (2) is a difficult object to compute,
while thermodynamics in Rindler space is relatively straightforward.

A few comments are in order about the generality of our proof.
No assumptions are made about the field theory, except that it must be
local and must possess Poincar\'e
invariance in the $x-t$ plane;
the theory need not be free.
The proof requires that the field be in its vacuum state,
and does not apply if some other
state is chosen.
It is also specific to the geometry
which we use; the equivalence is exact only for
spaces which are divided in half by a planar
boundary.   It gives a good approximation
to the density matrix which Srednicki considered$^1$, for which the
excluded region is a finite sphere, as long as the
 radius of the sphere is large.
Finite radius corrections are especially small for the
entropy of entanglement, which, as we will show,
comes predominantly from short-wavelength modes.
For example,
if the theory of a scalar field were modified
with a physical cutoff at the Planck
scale, making the entropy finite, then, for any sphere
of radius much larger than the Planck length,
the entropy of entanglement
outside the sphere would equal the thermal entropy
outside a black hole of the same radius.
Directly analyzing the entropy of entanglement in more
general geometries requires working with the definition (2);
for developments in this direction see the work of Callan and Wilczek$^4$.

To understand the physical basis for the equivalence of the two density
matrices, we focus attention on the quadrant $Q$ of Minkowski
space with $x > \vert t \vert$, since $Q$ is the part of spacetime
{\it causally disconnected} from the
$t=0, x<0$ half-space. Rindler and inertial observers have different ways of
restricting their measurements to $Q$.  A Rindler observer at a proper
distance $s$ from the horizon $x = \vert t \vert$ remains within
$Q$ by accelerating with a proper acceleration $1/s$; consequently
the observer
experiences a local temperature $1/2\pi s$.
An inertial observer, whose world line does not remain within $Q$, must
take a different approach. Consider an experiment performed by
an inertial observer at a distance $x_0$ from the imaginary
boundary at $x=0$.  From the moment the experiment
begins, its effects begin to
propagate from $x=x_0$ to $x=0$; given sufficient time they
will interact
with the $x<0$ half-space and return to $x=x_0$, allowing
the $x<0$ region to affect the
results of the experiment.  The experiment must therefore
be completed in a time $\Delta t < 2x_0$, which keeps it strictly within $Q$;
this leads to energy fluctuations of order $1/2x_0$, comparable to the
temperature experienced by Rindler observers at the proper distance $s=x_0$.
This highlights the
connection between ordinary quantum fluctuations and the
local temperature of Rindler space, and clarifies how the same density
matrix can describe both situations.

Since the density matrices are equivalent,
one may compute any observable
at $t=0$ either with the reduced density
matrix (1) or with thermodynamics in Rindler space; the results will
be identical.  It follows that the half-space entropy of entanglement
is formally equal to
the Rindler space entropy.  However, both entropies are divergent in most
quantum field
theories, and it seems unlikely that any observable exists which
would allow them to be directly compared.  Despite this, we will
consider, for a free scalar field, the finite thermal
entropy contained in a box in Rindler space.  This quantity
is adequate to study the universal
divergence and scaling properties
of the entropy of entanglement.  The coefficient of the divergence is not
universal, except in $1+1$ dimensions, and cannot be directly
compared with Srednicki's coefficient.

\bigskip
\goodbreak
\noindent{\bf ENTROPY CALCULATIONS}
\nobreak
\medskip
\nobreak
We now  present a few calculations of the Rindler
entropy in a box, following the original work of 't Hooft$^3$.
These results
will allow us, using the equivalence proved above,
to find the area dependence and degree of divergence of the
entropy of entanglement.   As we will see, naive scaling
analysis would have given the correct answer$^{1,2}$.
In more than two dimensions the
coefficient of the divergence is non-universal.

We begin in 1+1 dimensions.
 Let us compute at a fixed time the
entropy contained in a box with walls
at $x=x_0$  and $x=x_1$;\footnote{*}{$x_0$ and $x_1$ bound
the region over which a given experiment to measure the entropy
is sensitive.  They are {\it not} regulators
for the field theory; in fact we
perform the whole calculation in the continuum.}
the coordinate $x$ measures
proper distance from the Rindler horizon.
  For a well-defined counting
of the number of modes within the box boundary conditions must be chosen
at $x_0$ and $x_1$; some of our results are sensitive to this choice.
We work in the semiclassical (WKB) approximation; for a given (dimensionless)
Rindler energy $E$ the accumulated phase of the wave function is given by
$$\big(n(E) + \alpha\big) \pi = \int_{x_0}^{x_{\rm max}} {dx \over x}
                          \sqrt{E^2 - m^2 x^2} \, . \eqno(3)$$
This follows from adapting equation (3.7) of 't Hooft$^3$ for the
case of large black hole mass.  The upper limit of integration $x_{\rm max}$
is taken to be the smaller of the large distance cutoff $x_1$ and the point
$E/m$ at which the integrand vanishes.
The WKB quantization condition states that
for each $E$ such that $n(E)$ is an integer $j \geq 0$, there is a state
with energy $E_j = E$.
The constant $\alpha$ is to be
chosen according to the relevant boundary conditions.  If the upper limit of
integration is set by $x_1$ and the boundary conditions are
$\phi(x_0) = \phi(x_1) = 0$, then $\alpha = 1$; if the upper limit of
integration is set by $E/m$ then $\alpha = 3/4$ for $\phi(x_0) = 0$
and $\alpha = 1/4$ for $\phi'(x_0) = 0$.  The WKB approximation should be
quite accurate in this potential, even for low lying energy levels$^8$.

The free energy is given by a sum over energy eigenstates
$$\beta F = \sum_j \log\left(1 - e^{-\beta E_j}\right)\,;$$
the entropy in Rindler space is
$S = \left(\beta {\partial \over \partial \beta}
- 1\right)(\beta F)$ with $\beta$ set equal to $2 \pi$.
For small $\beta m x_0$ the sum over states in the partition function
is dominated by highly excited states, so we may introduce a smoothed
density of states $g(E) \equiv {dn(E) \over dE}$ and replace
$\sum_j \rightarrow \int dE \, g(E)$.  Note that we shall not see any
dependence on the choice of boundary conditions, as the density of states
is insensitive to the value of $\alpha$ in (3).  Integrating by parts gives
$$\beta F = - \beta \int dE \, {n(E) \over e^{\beta E} - 1}$$
with the convention $n(E) = 0$ for $E < m x_0$.
Expanding the Bose-Einstein distribution in powers of
$e^{-\beta E}$ and performing first the $E$ integration, then the $x$
integration, gives
$$\beta F = {1 \over \pi \beta} \sum_{n = 1}^\infty {1 \over n^2}
                \left[K_0(n \beta m x_1) - K_0(n \beta m x_0)\right]\,.
\eqno(4)$$
In appendix A we show that these sums behave as
$$\sum_{n=1}^\infty {1 \over n^2} K_0(n t)
       \sim -{\pi^2 \over 6} \log t + {\rm const.} \qquad t \ll 1 \,.$$
This gives the leading behavior of the free energy $\beta F$
in the regimes
$$\eqalign{
&x_0 \ll x_1 \ll {1 \over m} \, : \qquad
    - {1 \over \pi \beta} \, {\pi^2 \over 6} \log {x_1 \over x_0} +
              {\cal O}(mx_1)\cr
&x_0 \ll {1 \over m} \ll x_1 \, : \qquad
    - {1 \over \pi \beta} \left[{\pi^2 \over 6}\log {1 \over \beta m x_0}
                                           + {\cal O}(1)\right]\cr}$$
with corresponding entropies
$$\eqalign{
&x_0 \ll x_1 \ll {1 \over m} \, : \qquad
    {1 \over 6} \, \log {x_1 \over x_0} + {\cal O}(mx_1) \, ; \cr
&x_0 \ll {1 \over m} \ll x_1 \, : \qquad
    {1 \over 6} \, \log {1 \over m x_0} + {\cal O}(1) \,.\cr}
\eqno(5)$$
The first expression matches the exact $m = 0$ result obtained
in conformal field theory$^{4,5}$.  In the second,
intermediate mass case, we see that the mass takes over from $x_1$ in
setting a large distance cutoff.

We now consider the behavior in 1+1 dimensions for $\beta m x_0 > 1$.
Since $\beta E_j\geq \beta mx_0 $ we may set
$$\beta F \approx - \sum_j e^{- \beta E_j}\,.$$
In this case only
the lowest lying modes in the box are occupied, and it is inappropriate
to replace the sum over energy levels with an integral.
  The choice of boundary conditions is quite important now;
for $\phi(x_0) = 0$
$$\beta F = - e^{-\beta m x_0} \left(e^{-1.84\beta(mx_0)^{1/3}}
                                  +  e^{-3.24\beta(mx_0)^{1/3}}
                                  + \cdots \right) $$
while for $\phi'(x_0) = 0$
$$\beta F = - e^{-\beta m x_0} \left(e^{-0.89\beta(mx_0)^{1/3}}
                                  +  e^{-2.59\beta(mx_0)^{1/3}}
                                  + \cdots \right) \,.$$
These energy levels are found by solving the
WKB quantization condition (3) in the
limit of large $mx_0$.  As expected, the free energy is
exponentially suppressed.  If the sum over states
were incorrectly approximated with an integral over $E$, the calculation
would again lead to (4), which in the limit of large mass shows
the correct $e^{-\beta m x_0}$ suppression but misses entirely the
subleading $e^{- \beta(mx_0)^{1/3}}$ factors.  We will see this
issue is relevant for the higher dimensional case.

To discuss the entropy in higher dimensions, we use a box
extending from $x=x_0$ to $x=x_1$, with sides of length $L$
in each of the
$d_\perp$ transverse dimensions.  We consider massless fields; finite
mass effects are presented in appendix B.
The free energy $\beta F$ can be expressed as
a sum over transverse modes
labeled by their transverse momentum $k_\perp$.
Each mode is equivalent to a 1+1 dimensional scalar field
with an effective mass
$m_{\rm eff} = \vert k_\perp\vert,$ whose free energy
$\beta F_{1+1}(m_{\rm eff},x_0,x_1,\beta)$ we
have already computed.
$$\beta F =   \sum_{k_\perp}\
\beta F_{1+1}(\vert k_\perp\vert, x_0,x_1,\beta)$$
Because a long distance cutoff in the longitudinal
direction is no longer needed for a finite result,
we can set $x_1 = \infty$; in this limit,
by dimensional analysis, $F_{1+1}$ is a function
only of $\vert k_\perp\vert x_0$ and $\beta.$
Approximating the sum
as an integral $(L/2\pi)^{d_\perp}\int d^{d_\perp}k_\perp$
and changing variables to $\kappa=k_\perp x_0$,
we have
$$\beta F =  \left({L \over 2 \pi x_0}\right)^{d_\perp}
                \int d^{d_\perp}\kappa\
\beta F_{1+1}(\vert\kappa\vert,\beta)
\eqno(6)$$
This is of the form
$\beta F = C_{d_\perp}(\beta) \left({L\over x_0}\right)^{d_\perp}$,
where $C_{d_\perp}(\beta)$ is a
dimension dependent function of $\beta$.

We now check that no important $L$ dependence was hidden
when the sum over $k_\perp$ was replaced by an integral.
(By dimensional analysis this will also ensure
there is no hidden $x_0$ dependence.)
When $\vert\kappa\vert \equiv \vert k_\perp\vert x_0 \sim x_0/L$,
the integral
should really be treated as a sum, whose details will depend on
the boundary conditions on the walls of the transverse box.    We can
estimate the importance of this effect by computing the
contribution to the free energy from the part of the integration
region up to $\vert\kappa\vert = c x_0/L$, for $c$ of order one.
$$
 \left({L \over 2 \pi x_0}\right)^{d_\perp}
              \int_{\vert\kappa\vert \leq {c x_0\over L}} d^{d_\perp}\kappa \ \
{\pi \over 6 \beta}\ \log { \beta \vert\kappa\vert}
\ \ \sim \ \ \log(\beta c x_0/L )\,.
$$
Relative to the $(L/x_0)^{d_\perp}$ dependence of the leading term in (6),
this is negligible.
We therefore conclude that the straightforward scaling analysis is
correct$^{1,2}$: in any dimension $d_\perp>0$
both the free energy and the entropy are proportional to the
area $L^{d_\perp}$ of the transverse box, and
have a $d_\perp$-th order power-law divergence $1/x_0^{d_\perp}$
as the near edge of the box approaches the boundary ($x_0\rightarrow 0)$.

The coefficient
$$C_{d_\perp}(\beta) =  \left({1 \over 2 \pi }\right)^{d_\perp}
        \int d^{d_\perp}\kappa\ \beta F_{1+1}(\vert\kappa\vert,\beta)$$
gets its largest contribution from modes with $\vert\kappa\vert$
of order $1/\beta$.  The function
$\beta F_{1+1}(\vert\kappa\vert \equiv m_{\rm eff}x_0,\beta)$ increases
logarithmically  until $\vert\kappa\vert  \sim 1/\beta $,
then decreases exponentially;  additional powers of
$\vert\kappa\vert$ in the integration measure
push the main contribution to $C_{d_\perp}(\beta)$ out somewhat further.
As discussed above, for $\beta m_{\rm eff}x_0\geq 1$
it is a poor approximation to replace
the sum over states with an integral over a smoothed density
of states.   The result `t Hooft obtained$^3$ for $C_2(\beta)$
may not be reliable for this reason.   One can in principle determine
$C_{d_\perp}(\beta)$ exactly, for some choice
of boundary conditions, by solving the Schr\"odinger
equation to obtain the spectrum of states;
this probably cannot be done analytically.  However,
since a mode with large effective mass has a free energy which is
quite sensitive to the
boundary conditions at $x=x_0$, different choices for the
boundary conditions will lead to
different values of $C_{d_\perp}(\beta)$.
This implies that
$C_{d_\perp}(\beta)$ is not a universal quantity.

Even if the coefficient $C_{d_\perp}(\beta)$ were exactly known
for a particular set of boundary conditions, it would be impossible
to compare it quantitatively with Srednicki's results.
Srednicki computes the entropy
in a region $x_0<x<x_1$ with $x_0=a/2$,
where $a$ is the spacing of the longitudinal lattice which he introduces as
a regulator.  (By ``longitudinal'' we refer to the direction perpendicular
to the surface of his sphere.)  In the continuum limit
$a\rightarrow 0$, $x_0\rightarrow 0$ as well,
and the entropy in his region
becomes infinite.  This quantity allows him to correctly study the scaling
 and the divergence
structure of the entropy, but since his observable is not defined in the
continuum, our proof cannot be used to
relate his result directly to any finite computation
in Rindler space.  To see the connection with Rindler space one must define
a measurable quantity which is finite in the continuum limit;
the physical observable must be
separated from the regulator of the field theory.

\bigskip
\goodbreak
\noindent{\bf CONCLUSIONS}
\nobreak
\medskip
\nobreak
We briefly review our main points.

A simple proof demonstrates that the density matrix
of a quantum field theory obtained in the vacuum state
by tracing over half of space is
identical to the thermal density matrix of the field in Rindler space.
This holds for any local, renormalizable theory which is Poincar\'e
invariant in the $x-t$ plane.
We discuss the physical basis for this
equivalence.  It follows that the entropies studied by Srednicki$^1$ and
by 't Hooft$^3$ are formally equal.
Unfortunately, with few known exceptions$^6$,
these entropies are infinite, in which case it is probably impossible to
define a physically sensible and finite
observable which can be used to directly compare them.

Using the formal equivalence of the entropies, we argue that the
entropy of entanglement for a free scalar field
is logarithmically divergent in $1+1$ dimensions and
$d_\perp$-order divergent in $(d_\perp+1)+1$ dimensions.  The normalization
of the entropy is universal in $1+1$ dimensions and non-universal
otherwise.  We also show that for $d_\perp>0$ the entropy of entanglement
is proportional to the area of the boundary.

We conclude by noting an important possible
implication of the equivalence of
the two density matrices.
Both Srednicki and 't Hooft compute entropies associated
with one-loop effects of scalar fields in a background metric.  In the
presence of general relativity, there should be a tree-level contribution
to both entropies.  In other words, corresponding to the Bekenstein-Hawking
entropy of the horizon of an infinitely massive black hole,
there should be an entropy associated
with an imaginary planar boundary in perfectly flat space, with $S=(Area)/4$.
If this boundary entropy ought to be interpreted as a classical
entropy of entanglement, then it suggests that whatever the
fundamental generalization of Einstein's theory of gravity, it should be
non-local at short distances, even at the classical level.
In particular, Susskind and collaborators$^6$ have suggested that
in string theory this entropy is associated  with strings which
classically straddle the boundary.

While this manuscript was in preparation, results related to those presented
here appeared in refs.~4 and 6.

\bigskip
\goodbreak
\noindent{\bf ACKNOWLEDGEMENTS}
\nobreak
\medskip
\nobreak
We thank Tom Banks and Stephen Shenker for their many valuable contributions
to this paper, and Leonard Susskind and Frank Wilczek for
interesting discussions.

\bigskip
\goodbreak
\noindent{\bf APPENDIX A: SUMS}
\nobreak
\medskip
\nobreak

We consider sums of the form
$$f(t) = \sum_{n = 1}^\infty {1 \over n^2} \, K_0(nt)$$
where $K_0$ is a modified Bessel function.
To extract the behavior for small $t$ we calculate the Mellin
transformation$^9$
$$\eqalign{
F(\xi)&\equiv \int_0^\infty dt \, t^{\xi-1} f(t)\cr
            &= 2^{\xi-2} \zeta(2+\xi) \Big[\Gamma\left({\xi \over
2}\right)\Big]^2 \,.\cr}$$
Suppose $F(\xi)$ has poles at
$\xi_1,\xi_2,\ldots$, and write the principle part of the Laurent expansion
of about the $i^{th}$ pole
$$F(\xi) = {b_{1i} \over \xi - \xi_i} + {b_{2i} \over (\xi - \xi_i)^2}
             + \cdots \,. $$
Then asymptotically for small $t$
$$f(t) = \sum_i t^{- \xi_i} \left(b_{1i} + {1 \over 1!} b_{2i} (-\log t)
                  + {1 \over 2!} b_{3i} (-\log t)^2 + \cdots\right) \, .$$
The sum is dominated by the double pole
from the gamma functions at $\xi = 0$, with subleading behavior
from the simple pole in the zeta function at $\xi=-1$. One finds
$$f(t) \sim -{\pi^2 \over 6} \log t + {\pi^2 \over 6}(\log 2 - \gamma)
                   + \zeta'(2) + {\cal O}(t) \,.$$
For large $t$, $f(t)$ is exponentially suppressed.

\bigskip
\goodbreak
\noindent{\bf APPENDIX B: FINITE MASS}
\nobreak
\medskip
\nobreak

The free energy of a scalar field of mass $m$, $1/L \ll m \ll 1/x_0$, is
equal to the massless free energy (6) plus a correction
$$\beta F_{m\neq 0}-\beta F_{m=0} = \left({L \over 2 \pi x_0}\right)^{d_\perp}
   \int d^{d_\perp}\kappa
     \Big[\beta F_{1+1}\left(\sqrt{\kappa^2+m^2x_0^2},\beta\right)
            -\beta F_{1+1}(|\kappa|,\beta)\Big]  \,.$$
For $d_\perp\leq 2$ the main contribution comes from the infrared,
$|\kappa|\sim mx_0$, where the behavior of the integrand is known from
our 1+1 dimensional calculations.
$$
\beta F_{m\neq 0}-\beta F_{m=0} \approx
\left({L \over 2 \pi x_0}\right)^{d_\perp} {\pi\over 6\beta}
   \int d^{d_\perp}\kappa\ \log[1+(mx_0/\kappa)^2]$$
This gives the corrections
$$\eqalign{
& {\pi \over 12} {L m \over \beta}\ \ \ \ \ \ (d_\perp = 1);\cr
& - {1 \over 24} {L^2 m^2 \over \beta} \log mx_0 \ \ (d_\perp = 2)\,.\cr} $$
Since these corrections are finite and logarithmically divergent, respectively,
and come from a region of the integral where the integrand is insensitive
to boundary conditions, we conjecture that their coefficients are
universal.
For $d_\perp>2$ the integral is dominated by $|\kappa|\gg mx_0$, where
we may expand the integrand in powers of $m$.
$$\eqalign{\beta F_{m\neq 0}-\beta F_{m=0}  &\approx
\left({L \over 2 \pi x_0}\right)^{d_\perp}
   \int d^{d_\perp}\kappa\ (mx_0)^2 {d\over d\kappa^2}
\Big[\beta F_{1+1}(|\kappa|,\beta)\Big] \cr
 &=- {(d_\perp-2)\pi^{d_\perp/2}\over\Gamma(d_\perp/2)}
(mx_0)^2\left({L \over 2 \pi x_0}\right)^{d_\perp}
   \int d^{d_\perp}\kappa\ \kappa^{d_\perp-3}
\Big[\beta F_{1+1}(|\kappa|,\beta)\Big]\cr
 &= -{1\over 4\pi} (mL)^2 \beta F|_{(d_\perp-2)} \cr}$$
This is proportional to the free energy
in $d_\perp-2$ dimensions, which is power-law divergent as
$x_0\rightarrow 0$ and depends on boundary conditions,
and so this correction is non-universal; still this result
is rather interesting.

\bigskip
\goodbreak
\noindent{\bf REFERENCES}
\nobreak
\medskip
\nobreak
\item{1.}  M.~Srednicki, {\it Phys.~Rev.~Lett.} {\bf 71}, 666 (1993).
\item{2.}  L.~Bombelli, R.~K.~Koul,
J.~Lee, and R.~D.~Sorkin, {\it Phys.~Rev.} {\bf D34}, 373 (1986).
See in particular footnote 17.
\item{3.}  G.~'t Hooft, {\it Nucl.~Phys.} {\bf B256}, 727 (1985).
\item{4.}  C.~Callan and F.~Wilczek, {\it On Geometric Entropy},
preprint IASSNS--HEP--93/87, hep-th/9401072 (January 1994).
\item{5.}  C.~Holzhey, Princeton University thesis (unpublished, 1993);
C.~Holzhey, F.~Larsen, and F.~Wilczek, {\it Geometric Entropy
in Conformal Field Theory}, preprint IASSNS--HEP 93/88.
\item{6.}  L.~Susskind and J.~Uglum, {\it Black Hole Entropy in Canonical
Quantum Gravity and Superstring Theory}, Stanford preprint SU--TP--94--1,
hep-th/9401070 (January 1994); L.~Susskind, {\it Some Speculations about
Black Hole Entropy in String Theory}, preprint RU--93--44, hep-th/9309145
(September 1993).
\item{7.}  The quantum theories resulting from the different choices of time
slicing will be identical.  D.~G.~Boulware, {\it Phys.~Rev.} {\bf D11}, 1404
(1975).
\item{8.}  J.~J.~Sakurai, {\it Modern Quantum Mechanics} (Addison--Wesley,
1985), p.~108.
\item{9.}  F.~Oberhettinger, {\it Tables of Mellin Transforms}
(Springer--Verlag, 1974), p.~7.
\vfill
\end